\documentclass[a4paper,10pt,aps,prd,showpacs,preprintnumbers,twocolumn,nofootinbib,superscriptaddress]{revtex4-2}

\usepackage[english]{babel}
\usepackage[utf8]{inputenc}
\usepackage[T1]{fontenc}
\usepackage{newtxtext,newtxmath}
\usepackage{color}
\usepackage{url}
\usepackage{hyperref}
\usepackage[dvipsnames]{xcolor}
\usepackage{graphicx}
\usepackage{float}
\usepackage{amsmath}
\usepackage{mathrsfs}
\usepackage{verbatim}
\usepackage{booktabs}
\usepackage{multirow}
\usepackage{lettrine}
\usepackage{natbib}
\usepackage{empheq} 

\def \be {\begin{equation}}
\def \ee {\end{equation}}
\def \dd {\mathrm{d}} 
\def \t {\tilde}
\def \p {\partial}
\def \l {\left}
\def \r {\right}
\def \bs {\boldsymbol}
\def \dd {\mathrm{d}} 
\def \bs {\boldsymbol}

\newcommand{\e}[1]{_{\rm #1}}

\usepackage{soul}

\definecolor{citeco}{HTML}{2E3093}
\definecolor{linksco}{HTML}{2E3093}

\begin{document}

\title{Multimessenger lensing time delay
as \\ a probe of the graviton mass}

\author{Elena Colangeli}
\email{elena.colangeli@port.ac.uk}
\affiliation{Institute of Cosmology and Gravitation, University of Portsmouth, \\
Burnaby Road, Portsmouth PO1 3FX, United Kingdom}
\author{Charles Dalang}
\email{charles.dalang@phys.ens.fr}
\affiliation{Institute of Cosmology and Gravitation, University of Portsmouth, \\
Burnaby Road, Portsmouth PO1 3FX, United Kingdom}
\affiliation{Queen Mary University of London, \\
Mile End Road, London E1 4NS, United Kingdom}
\affiliation{Institut Philippe Meyer, Département de Physique, École Normale Supérieure (ENS), Université PSL,
Paris, France}
\author{Tessa Baker}
\email{tessa.baker@port.ac.uk}
\affiliation{Institute of Cosmology and Gravitation, University of Portsmouth, \\
Burnaby Road, Portsmouth PO1 3FX, United Kingdom}

\begin{abstract}
Gravitational lensing is a powerful probe of cosmology and astrophysics. With the prospect of the first strongly lensed gravitational waves on the horizon, we highlight an opportunity to test fundamental physics. In this work, we assume a nonzero mass for the graviton, which leads to gravitational waves following timelike geodesics instead of null geodesics. We derive standard gravitational lensing equations, such as the scattering angle, the time delay between different images and the magnification, which normally rely on the assumption of null geodesics.
We show that a single strongly lensed multimessenger event is enough to constrain the graviton mass to $m< 3 \cdot 10^{-23}$eV/c$^{2}$. Notably this constraint is independent of the lens model, of the waveform model, and of cosmology. Additionally, we explore magnification of images and find that they offer at least 3 orders of magnitude weaker bounds than the time delay, and have a dependence on the correct modeling of the lens and cosmology.
\end{abstract}

\maketitle

\section{Introduction} 

\noindent Strong gravitational lensing of electromagnetic (EM) waves is a well-established probe of cosmology \cite{TDCOSMO:2025dmr}. In recent years gravitational wave (GW) strong lensing has been actively developing both theoretically and observationally \cite{Wang:1996as, Takahashi:2003ix, Smith:2017jdz, Cremonese:2018cyg, Li:2018prc, Hannuksela:2019kle, Ezquiaga:2020gdt, Ezquiaga:2020spg, Bulashenko:2021fes, Wierda:2021upe, Tambalo:2022plm, LIGOScientific:2023bwz} as a complementary method to EM lensing -- though searches have yet to find conclusive lensed signals. This is due to the rarity of these events combined with current detector sensitivity limitations \cite{LIGOScientific:2023bwz}; we expect to observe one lensed event for every $\sim 1500$ unlensed detections \cite{Wierda:2021upe}.\footnote{To contextualize this, approximately 200 unlensed events have been reported so far.} The combination of EM and GW gravitational lensing naturally leads to the possibility of \textit{multimessenger lensing} \cite{Smith:2025axx}: a multimessenger strongly lensed event, which is often called a ``golden'' scenario. Despite the predicted scarcity of strongly lensed GW signals, if one is observed it will be straightforward to find the lensed host galaxy, thanks to new wide-field observatories such as Euclid \cite{Euclid:2024yrr} and LSST \cite{LSST:2008ijt}. Prospects are quite optimistic due to the sheer number of strongly lensed galaxies: the recent first data release from the Euclid mission \cite{aussel:2025} reported 497 new strong lenses, and forecasts predict a total of $\sim 75,000$ lenses by the end of the survey \cite{lines2025euclid}. Additionally, LSST forecasts $\sim 70,000$ strong lenses (for a conservative estimate) in 10 years \cite{Shajib:2024yft}, with more optimistic predictions up to order $\sim 400,000$ strong lenses from Euclid, LSST, and DES combined \cite{Collett:2015roa}. Detection of a golden event would offer the opportunity to perform tests of general relativity in addition to cosmological constraints. 

Modified gravity models have been explored in the context of unlensed multimessenger signals \cite{Amendola:2017ovw,LISACosmologyWorkingGroup:2019mwx,Belgacem:2018lbp,Lagos:2019kds,Baker:2017hug,Dalang:2019rke,Dalang:2020eaj,Baker:2020apq,Colangeli:2025bnb,Ezquiaga:2018btd}. 
In the context of gravitational wave lensing, a few studies of modified gravity have been performed. For example, different authors investigated tests of local Lorentz violations \cite{Biesiada:2007rk}, anomalous GW damping \cite{Finke:2021znb,Narola:2023viz} anomalous speed of GWs \cite{Collett:2016dey,Fan:2016swi} or both \cite{Ezquiaga:2020dao}. The effect of Horndeski gravity on a lens was studied in \cite{Kumar:2021cyl} and gravitational lensing of electromagnetic signals was investigated for a subset of Horndeski theories in which GWs travel at light speed in \cite{Bessa:2023ykd}. The frequency-dependent amplification factor of lensed gravitational waves was proposed as a novel test of the graviton mass in \cite{Chung:2021rcu}.

In this work, we bridge the gap between lensing in massive gravity and multimessenger gravitational wave events. We perform a comprehensive theoretical calculation for a golden event in the case of a massive graviton. We investigate how the presence of a mass in the dispersion relation of gravitational waves affects geodesics, time delays, and magnification in a gravitational lensing context. We outline observational prospects of a golden event in this scenario. We show how comparing EM and GW signals allows one to impose model-independent bounds on the graviton mass, adding to the existing dispersion relation tests of gravity, which find $m \leq 1.27 \cdot 10^{-23}$eV/c$^2$ (90\% confidence level) \cite{LIGOScientific:2021sio}.

We show that a fully model-independent bound on the graviton mass may be imposed from the time delay of a strongly lensed  multimessenger event, which will provide an alternative constraint, complementary to both GW dispersion constraints mentioned above and non-GW bounds from the Solar System, clusters, binary pulsars \cite{Poddar:2021yjd} and weak lensing (see \cite{deRham:2016nuf} for a comprehensive review). 

This article is structured as follows. In Sec.\,\ref{sec:massive_geodesics}, we show how a mass may affect the dispersion relation of gravitational waves and derive the geodesic equation. In Sec.\,\ref{sec:scattering_angle}, we show how the scattering angle of a lensed GW is affected by the mass of the graviton. In Sec.\,\ref{sec:time_delay_constraint}, we compute the time delay between different images for a strongly lensed massive gravitational wave and explore multimessenger constraints. In Sec.\,\ref{sec:magnification_constraint}, we derive the magnification of massive GWs from scratch and show that the constraints that can be obtained on the graviton mass from the comparison of EM versus GW magnification is weaker than from the time delay. 
Finally, we conclude in Sec.\,\ref{sec:conclusion}. We chose the $(-,+,+,+)$ metric signature and units are such that $c=1 = \hbar$.

\section{Massive geodesics}\label{sec:massive_geodesics}

\noindent In this section, we make a pedagogical introduction on how a mass term may alter the geodesics of gravitational waves. We study this phenomenologically and defer more formal aspects of massive gravity to the literature \cite{deRham:2014zqa} and references therein. Note that a UV completion of Lorentz violating massive gravity was proposed in \cite{Blas:2014ira}.
Suppose that the equation of motion for the metric perturbation reads 
\begin{align}
\label{eq:linear_EFE}
    (\Box - m^2) h_{\mu \nu} + 2 R_{\mu\rho \nu \sigma} h^{\rho \sigma} \simeq 0 \,,
\end{align}
where we split the metric $\bar{g}_{\mu\nu} = g_{\mu\nu} + h_{\mu\nu}$ into a background spacetime described by $g_{\mu\nu}$ and gravitational waves, described by $h_{\mu\nu}$. In Eq.\,\eqref{eq:linear_EFE}, the box operator and the Riemann tensor $R_{\mu\rho\nu\sigma}$ are defined with respect to the background metric $g_{\mu\nu}$, i.e.\,$\Box \equiv g^{\mu\nu} \nabla_\mu \nabla_\nu$ and $m$ denotes the graviton mass. Such an equation of motion may arise from the variation of the following action for a massive graviton (see for example \cite{Bernard:2017tcg})
\begin{align}
S[h_{\mu\nu}] & = -M\e{Pl}^2\int \dd^4 x \sqrt{-g} \Big[ h_{\mu\nu} \mathcal{D}^{\mu\nu \rho \sigma} h_{\rho \sigma} \label{eq:Action} \\
& \quad - \Lambda \l( h_{\mu\nu} h^{\mu\nu} - \frac{1}{2}h^2\r) + \frac{m^2}{2}\l( h_{\mu\nu} h^{\mu\nu} - \frac{1}{2}h^2\r) \Big] \nonumber
\end{align}
where $M\e{Pl}$ denotes the Planck mass, $\Lambda$ denotes the cosmological constant, $h= g^{\mu\nu}h_{\mu\nu}$ and the operator
\begin{align}
\mathcal{D}^{\mu\nu \rho \sigma} h_{\rho \sigma} & \equiv -\frac{1}{2}\Big[ g^{\mu\rho} g^{\nu\sigma} \Box + g^{\rho \sigma}\nabla^\mu \nabla^\nu - g^{\mu \rho} \nabla^\sigma \nabla^\nu \\
& \quad - g^{\nu \rho}\nabla^\sigma \nabla ^\mu - g^{\mu\nu} g^{\rho \sigma}\Box + g^{\mu \nu} \nabla^\rho \nabla^\sigma \Big] h_{\rho \sigma} \,.\nonumber
\end{align}
The equation of motion associated with the action in \eqref{eq:Action}, Eq.\,\eqref{eq:linear_EFE} follows with the constraints $\nabla^\mu h_{\mu\nu} \simeq 0$, $h\simeq 0$, if the background spacetime satisfies $R_{\mu\nu} = \Lambda g_{\mu\nu}$. While there exists this formal possibility, the relevant aspect for our work is the fact that the metric perturbation factors out of the $(\Box - m^2)$ term in Eq.\,\eqref{eq:linear_EFE}. Note that a more complicated structure arises for de Rham-Gabadadze-Tolley (dRGT) \cite{deRham:2010kj} ghost-free massive gravities \cite{Bernard:2015mkk,Bernard:2014bfa}. 

Adopting Eq.\,\eqref{eq:linear_EFE} as the equation of motion, we make the following wave Ansatz
\begin{align}
\label{eq:h_ansatz}
    h_{\mu \nu}(x)=  H_{\mu \nu}(x) e^{\mathrm{i} \varphi(x)}\,,
\end{align}
where $H_{\mu \nu}$ describes the real amplitude and polarization of the wave, while $\varphi$ describes the phase. In principle, $H_{\mu\nu}$ can describe up to five polarization modes, which follows from the 10 degrees of freedom of a symmetric rank 2 tensor deducted from the 5 constraints $\nabla^\mu h_{\mu\nu} \simeq 0 \simeq h$. The wave vector is defined as $k_\mu \equiv \nabla_\mu \varphi$. Plugging Eq.\,\eqref{eq:h_ansatz} into Eq.\,\eqref{eq:linear_EFE}, we get
\begin{align}
   \nonumber - (k^\alpha k_\alpha + m^2)H_{\mu \nu} & + \l(\Box H_{\mu \nu} -2H^{\alpha \beta} R_{\mu \alpha \nu \beta}\r) 
   %\label{eq:real_part}
   \\
    &+  \mathrm{i}(2 k^\alpha\nabla_\alpha + \nabla_\alpha k^\alpha)H_{\mu \nu} = 0 \, \label{eq:imaginary_part}
\end{align}
The real part and imaginary part of this equation should vanish independently. We work in the geometric optics approximation, where the phase varies much faster than the amplitude of the wave and the typical scale over which the background spacetime varies.\footnote{Note that if this assumption is broken, the time delay between GWs and electromagnetic signals may contain additional corrections \cite{Takahashi:2016jom}.} In this scenario, the first term dominates over the last two\footnote{Note that these terms can lead to weak polarization distortions of lensed gravitational waves in certain configurations \cite{Cusin:2019rmt,Dalang:2021qhu}.}
in the first line of Eq.\,\eqref{eq:imaginary_part} such that we are left with
\begin{align}
k^\mu k_\mu = -m^2 \,,\label{eq:kmu_kmu}
\end{align}
which describes the dispersion relation of gravitational waves. While we have presented an example action that leads to Eq.\,\eqref{eq:kmu_kmu} for pedagogical purposes, we consider this dispersion relation Eq.\,\eqref{eq:kmu_kmu} to be the starting assumption of this work. Since we have chosen the phase $\varphi$ to be polarization independent, Eq.\,\eqref{eq:kmu_kmu} applies to any polarization.  
The imaginary part [Eq.\,\eqref{eq:imaginary_part}] can be integrated to solve for the amplitude and polarization of the gravitational wave and to show that the polarization is parallel transported along the geodesic to leading order in geometric optics (see for example \cite{Dalang:2021qhu}).

One can derive the geodesic equation by taking the covariant derivative of Eq.\,\eqref{eq:kmu_kmu} to find
\begin{align}
\nonumber 0 & = k^\mu (\nabla_\nu k_\mu) = k^\mu (\p_\nu \p_\mu \varphi -\Gamma^\lambda_{\nu\mu} k_\lambda) \\
\nonumber& = k^\mu (\p_\mu \p_\nu \varphi - \Gamma^\lambda_{\mu\nu} k_\lambda)\\
& = k^\mu \nabla_\mu k_\nu \, ,\label{eq:geodesic_equation}
\end{align}
where we have used that $k_\mu$ is the gradient of the phase and the Christoffel symbols are symmetric in the lower two indices. Using the definition of $k^\mu = \dd x^\mu/ \dd \lambda$ in terms of an affine parameter $\lambda$, this expression can be integrated to find
\begin{align}
k^\nu(\lambda_o) = k^\nu(\lambda_s) - \int_{\lambda_s}^{\lambda_o} \dd \lambda \,\Gamma^\nu_{\mu\lambda} k^\mu k^\lambda \,,\label{eq:integrated_geodesic}
\end{align}
where subscripts $o$ and $s$ refer to the position of the observer and the source, respectively. From this expression, it seems that given the same initial direction of propagation, a massive and massless graviton will generally travel along different curves since the background Christoffel symbols contracts with all components of $k_\mu$, which satisfy different constraints: the massive case follows Eq.\,\eqref{eq:kmu_kmu}, while in the case of a massless graviton, we would have $k^\mu k_\mu = 0$. In the next section, we illustrate the different scattering angles for massive and massless gravitons in a simple setting of a pointlike lens.

\section{Scattering angle}\label{sec:scattering_angle}

\noindent To illustrate the fact that the curves described by a massive and massless geodesic differ in the weak-field and low-mass limit, we assume the lens to be well described by a weak-field Schwarzschild metric in isotropic coordinates
\begin{align}
\label{eq:Weak_field_metric}
\dd s^2 = -(1+2 U)\dd t^2 + (1-2 U)\dd \bs{x}^2\,,
\end{align}
where $U=-R_s/(2||\bs{x}||)$ is the gravitational potential such that $|U|\ll 1$, $||\bs{x}|| = \sqrt{x^2 + y^2 + z^2}$ is the distance from the lens, which is centered at the origin of the coordinate system, and $R_s$ is its Schwarzschild radius. To first order in the metric potential, the nonvanishing Christoffel symbols read
\begin{align}
\Gamma^i_{00} & = \p^i U\,, \\
\Gamma^0_{i0} & = \p_i U \,, \\
\Gamma^i_{jk} & = -\delta^i_j \p_k U  - \delta^i_k \p_j U  +\delta_{jk} \p^i U \,.
\end{align}
We can solve Eq.\,\eqref{eq:integrated_geodesic} perturbatively in the potential. We define $k^\mu(\lambda) = \bar{k}^\mu(\lambda)+\delta k^\mu(\lambda)$ such that $\bar{k}^\mu(\lambda) = \mathcal{O}(U^0)$ and $\delta k^\mu(\lambda) = \mathcal{O}(U^1)$. In this case, the leading order contribution $\bar{k}^i(\lambda) = k^i(\lambda_s)$ represents the undeflected wave vector, which is identical all along the geodesic. Here, we express it in terms of the initial condition $k^i(\lambda_s)$. 

It is convenient to decompose the wave vector on a tetrad basis, formed by a timelike vector $u^\mu$, which can describe the four-velocity of an observer. We also define an orthogonal spacelike vector $d^\mu$ such that $d^\mu u_\mu =0$. Both of these vectors are normalized such that $u^\mu u_\mu = -1$ and $d^\mu d_\mu = 1$. The vector $d^\mu$ can be thought as the spatial direction of propagation of the wave, orthogonal to which one could build a screen basis, for example the Sachs basis. The wave vector expanded on that basis reads 
\begin{align}
k^\mu = \omega u^\mu + k d^\mu\,.
\end{align}
One can easily check that $k^\mu k_\mu = -m^2$ with $\omega^2 = k^2 + m^2$. The meaning of the affine parameter can be understood by taking the projection $d_\mu$ on an infinitesimal coordinate displacement $\dd x^\mu$. In the frame of the observer, the distance traveled by a photon corresponds to 
\begin{align}
\dd \ell = d_\mu \dd x^\mu = d_\mu k^\mu \dd \lambda = k \dd \lambda\,.\label{eq:affine_parameter}
\end{align}
Choosing an orientation, this wave vector can be written as  $\bar{k}^\mu(\lambda) = (\omega,0,0,k)$, which describes the geodesic of a massive graviton traveling in Minkowski space toward the $+\bs{\hat{e}_z}$ direction. If the wave is sent from $\bs{x} =(x=b, y=0, z\to -\infty)$, where $b$ is the impact parameter, we expect the wave to be deflected toward the $-\bs{\hat{e}_x}$ direction. 

For a wave propagating toward the $+\bs{\hat{e}_z}$ direction, this implies that $\dd z = k \dd \lambda$, as per Eq.\,\eqref{eq:affine_parameter}. We now compute the integrals which describe $\delta k^i$. To first order in $U\ll 1$, Eq.\,\eqref{eq:integrated_geodesic} leads to 
\begin{align}
\delta k^i(\lambda_o) = -\int_{\mathbb{R}} \frac{\dd z}{k} \l( \Gamma^i_{00} (\bar{k}^0)^2 + \Gamma^i_{jk} \bar{k}^j\bar{k}^k \r) \,.
\end{align}
One can easily show that $\delta k^z(\lambda_o) = 0 = \delta k^y(\lambda_o)$. However, along the $\bs{\hat{e}}_x$ direction,
\begin{align}
\delta k^x(\lambda_o) & = -\frac{\omega^2 + k^2}{k} \frac{R_s}{b} \label{eq:delta_kx_omega_k} = - \frac{2 k R_s}{b} \l( 1 + \frac{m^2}{2k^2}\r)
\,. 
\end{align}
Therefore, the normalized deflected wave vector $\bs{\hat{k}}$ at the observer, which we denote with a hat, reads
\begin{align}
\bs{\hat{k}} \simeq \frac{1}{\l(1+\frac{2 R_s^2}{b^2} \l(1+ \frac{m^2}{k^2} \r) \r)}\l( -\frac{2R_s}{b}\l(1+ \frac{m^2}{2 k^2} \r),0,1\r) \,,
\end{align}
up to $\mathcal{O}(R_sm^4/(b\omega^4) )$. The scattering angle reads
\begin{align}
\hat{\alpha} = \arccos \l( \bs{\hat{k}} \cdot \bs{\hat{\bar{k}}}\r)  = \frac{2 R_s}{b} \l( 1+ \frac{m^2}{2\omega^2} \r)\,, 
\end{align}
where the $m^2/\omega^2$ holds the leading order difference between the deflection angle of a massive and a massless geodesic. Massive gravitons are more deflected than massless particles. Their images form at 
\begin{align}
\bs{\theta}\e{g}^{\rm I} \simeq \bs{\theta}_\gamma^{\rm I}\l( 1+ \frac{m^2}{2\omega^2}\r)\,,\label{eq:images_GW_vs_EM}
\end{align}
where $\bs{\theta}_\gamma^{\rm I}$ denotes the I'th image position of their massless counterpart. Even if gravitational waves have poor sky localization ($\sim 0.01$ deg$^2$ in the most optimistic scenarios \cite{Iacovelli:2022bbs}) compared to electromagnetic signals which can be located to sub-arcsec precision, this difference may be quite important when comparing electromagnetic and GW time delay to constrain the graviton mass. Note that this different deflection angle with respect to massless gravitons also holds for GWs traveling at different speeds than the speed of light. In fact, if $\omega \neq k$, the deflection angle is affected, as may be understood from Eq.\,\eqref{eq:delta_kx_omega_k}. We depict the different scattering angles in Fig.\,\ref{fig:Massive_graviton_lensing}.

In the next section, we compute the time delay between different images of a strongly lensed system. Having determined that massive geodesics follow different geodesics than their massless counterpart, we compute corrections to the time delay due to the massive geodesics followed by GWs to order $\mathcal{O}(m^2/\omega^2)$.

\section{Time-delay constraint}\label{sec:time_delay_constraint}

\noindent In this section, we start from the massive geodesic equation and derive the time delay between two images strongly lensed by a foreground lens in the context of a FLRW (Friedmann–Lemaître–Robertson–Walker) background spacetime. We first compute the geometrical time delay in Sec.\,\ref{sec:geometric_time-delay} and then focus on the Shapiro time delay in Sec.\,\ref{sec:shapiro_time-delay}. The total time delay is the sum of these two contributions, as detailed in Sec.\,\ref{sec:total_time-delay}.
\begin{figure}
    \centering
    \includegraphics[width=0.6\columnwidth]{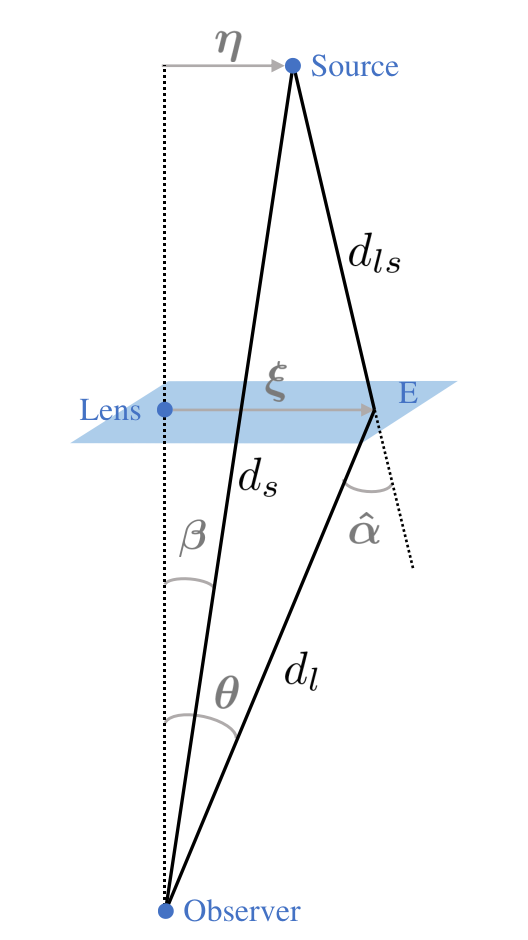}
    \caption{Schematic of the lensing configuration: a signal traveling from a source and encountering a lens travels along a deflected path $d_{ls}+d_{l}$ before reaching the observer. Angular positions of the source $\bs{\beta}$ and image $\bs{\theta}$ are shown along with the scattering angle
    $\bs{\hat{\alpha}}$. Vectors $\bs{\eta}$ and $\bs{\xi}$ %
    represent the physical positions on the source and lens plane respectively, where we highlight the lens plane $E$. The distance $d_s$ is the undeflected path the signal would take in the absence of the lens.}
    \label{fig:lensing_configuration}
\end{figure}

\subsection{Geometric time delay}\label{sec:geometric_time-delay}

\noindent We consider a flat FLRW metric in conformal time, which is described by the following line element:
\begin{align}
\dd s^2 = g_{\mu\nu} \dd x^\mu \dd x^\nu = a^2(\eta) \left[ -c^2 \dd\eta^2 + \delta_{ij} \dd x^i \dd x^j \right]\,. 
\label{eq:metric_FLRW}
\end{align}
Applying Eq.\,\eqref{eq:kmu_kmu} to this background spacetime, one finds that a massive GW trajectory would obey 
\begin{align}
- a^2(\eta) \left( \frac{\dd \eta}{\dd \lambda}\right)^2 + a^2(\eta) \delta_{ij} \frac{\dd x^i}{\dd \lambda} \frac{\dd x^j}{\dd \lambda} = - m^2\, .
\end{align}
This can be rewritten as
\begin{align}
    \frac{\dd \eta}{\dd \lambda} = \sqrt{\delta_{ij} \frac{\dd x^i}{\dd \lambda} \frac{\dd x^j}{\dd \lambda} +  \frac{m^2}{a^2} } 
    &\simeq ||\bs{k}||\left(1 + \frac{m^2}{2a^2||\bs{k}||^2}\right) \, ,
\end{align}
where $||\bs{k}|| \equiv \sqrt{\delta_{ij} \frac{\dd x^i}{\dd \lambda} \frac{\dd x^j}{\dd \lambda}}$ is the norm of the spatial component of the wave vector. In the last step, we assume that $m^2\ll a^2||\bs{k}||^2$.\footnote{Note that $m^2/||\bs{k}||^2 \simeq m^2/\omega^2$ to leading order.} Integrating both sides of this expression along the affine parameter yields:
\begin{align}
\label{eq:Delta eta}
    \Delta \eta = \int_{\lambda_1}^{\lambda_2} \dd \lambda  ||\bs{k}|| + \int_{\lambda_1}^{\lambda_2} \dd \lambda \frac{m^2}{2a^2||\bs{k}||} \, .
\end{align}
The first integral represents the comoving distance between $\lambda_1$ and $\lambda_2$, while the second integral is a mass correction term. The geometrical time delay between two signals emitted simultaneously and traveling the deflected versus the undeflected path reads
\begin{align}
\label{eq:delta eta d}
\Delta \eta\e{geo} = d_l + d_{ls} - d_{s} +  \int_{\lambda_o}^{\lambda_o + \Delta \lambda_o} \dd \lambda \frac{m^2}{2a^2||\bs{k}||} \, .
\end{align}
where $\Delta \lambda_o$ represents the difference in the affine parameter required to travel the deflected path with respect to the undeflected path and $d_l$, $d_s$ and $d_{ls}$ are the comoving distances which are depicted in Fig.\,\ref{fig:lensing_configuration}. For a time delay which is much smaller than the Hubble time, the scale factor can be considered to be constant over the time interval spanned by $\Delta \lambda$, i.e.\,$a(\eta_o) =a_0=1$. This also implies that $\Delta \eta\e{geo} = \Delta t\e{geo}$. Solving the geodesic equation [Eq.\,\eqref{eq:geodesic_equation}] yields $a^2||\bs{k}||=\text{const}$, which implies that the second integral reads
\begin{align}
\nonumber\Delta t\e{geo} & =  (d_l + d_{ls} - d_{s}) + \frac{m^2}{2||\bs{k}||^2}\int_{\lambda_o}^{\lambda_o + \Delta \lambda_o} \dd \lambda ||\bs{k}||\\
& =(d_l + d_{ls} - d_{s}) \l(1 + \frac{m^2}{2\omega^2}\r)\,.
\end{align}
In the above, we used the fact that the integral in the first line is the same as the first term in Eq.\eqref{eq:Delta eta}, translating the integral to the difference in distances. The correction factor multiplies the null geodesic result for which some well-known trigonometry can be applied to arrive at \cite{Schneider:1992}
\begin{align}
\Delta t\e{geo} & = \frac{d_{ls}d_l}{2d_s} \bs{\hat{\alpha}}^2 \l(1 + \frac{m^2}{2\omega^2}\r)\,.
\end{align}
Next, we can use the relation between the comoving distance and the angular diameter distance $D_s = d_s/(1+z_s)$, $D_l = d_l/(1+z_l)$, $D_{ls} = d_{ls}/(1+z_s)$ and $D_{\ell s} \bs{\hat{\alpha}} = D_s (\bs{\theta}- \bs{\beta})$, which implies that we can rewrite the geometric time delay as
\begin{align}
\Delta t\e{geo} = (1+z_l) \frac{D_s D_\ell}{D_{\ell s}}  (\bs{\theta}- \bs{\beta})^2 \l(1 + \frac{m^2}{2\omega^2}\r) \, ,\label{eq:geom_time_delay}
\end{align}
where $z_l$ is the redshift of the lens, $D_s$, $D_l$  and $D_{ls}$ are the angular diameter distances to the source, to the lens and between the lens and the source, while $\bs{\theta}$ and $\bs{\beta}$ are the angular positions of the image and source respectively. The formula also holds for a nonflat FLRW geometry \cite{Schneider:1992}.

\subsection{Shapiro time delay} \label{sec:shapiro_time-delay}

\noindent In this section, we compute the second contribution of the time delay, the Shapiro time delay. For this purpose, we assume that the metric is well described by the line element in Eq.\,\eqref{eq:Weak_field_metric}, with the static weak-field gravitational potential $U(\bs{x})\ll 1 $. Note that we use a different line element for this contribution, as we take into account the effect of the gravitational potential of the lens, for which cosmology is irrelevant. We rearrange Eq.\,\eqref{eq:kmu_kmu} into 
\begin{align}
   \nonumber k^0 \equiv \frac{\dd t}{\dd \lambda} &= \pm \sqrt{
    -\frac{g_{ij}}{g_{00}} \frac{\dd x^i}{\dd \lambda}\frac{\dd x^j}{\dd \lambda} 
    -\frac{m^2}{g_{00}}}
    \\ &\simeq \pm ||\bs{k}||\l(1 - 2U\r)\l(1 + \frac{m^2}{2\omega^2}\r)  + \mathcal{O}\l(\frac{m^4}{\omega^4}, U^2\r) \,, \label{eq:k0}
\end{align}
where we replaced the metric \eqref{eq:Weak_field_metric}, used the definition of $\bs{k}$ and the fact that $U\ll 1$, and $m^2\ll \omega^2 \simeq ||\bs{k}||^2$. We also neglect second order terms, i.e.\,$\mathcal{O}(m^4/\omega^4, U^2)$. Integrating along $\dd \lambda$ between the source and the observer, one finds the positive time elapsed between the two events to be
\begin{align}
    (t_{o} - t_{s}) &= \l( 1+\frac{m^2}{2\omega^2}\r) \int_{\lambda_s}^{\lambda_o} \left[1 - 2U(\gamma(\lambda)) \right] ||\bs{k}||  \dd \lambda \nonumber
    \\ &= \l( 1+\frac{m^2}{2\omega^2}\r) \l( d  -2  \int_{\lambda_s}^{\lambda_o} U(\ell)\dd \ell \r) \, , \label{eq:Integral_Shapiro}
\end{align}
where the first term in the second bracket is the path length along the affine parameter, the contribution to the time delay of which we have calculated in detail in the previous section [see Sec. \ref{sec:geometric_time-delay}, Eq.\,\eqref{eq:geom_time_delay}]. The second term in the second bracket is the Shapiro time delay, which arises because of the presence of the gravitational field \cite{Schneider:1992}. This contribution depends on the path $\gamma(\lambda)$. The graviton mass acts as a correction to the Shapiro time delay. 

To compute the Shapiro time delay (which we denote $\Delta t\e{sha}$), one can exploit the fact that for a point mass located at the origin, $U(\bs{x}) = -G M/||\bs{x}||$. Since the result for the time delay is linear in the mass, it is possible to express this integral as
\begin{align}
-2  \int_{\lambda_s}^{\lambda_o} U(\ell)\dd \ell = - 4 G \int_{\mathbb{R}^2} \dd^2 \bs{\xi'} \Sigma(\bs{\xi'})\log\l( ||\bs{\xi}-\bs{\xi'}||\r) + \hbox{const}\,,
\end{align}
where $\Sigma(\bs{\xi})$ is the surface mass distribution, described in terms of a coordinate $\bs{\xi}$ in the lens plane. Since this time delay accumulates at the lens, the appropriate lens redshift factor should be included leading to 
\begin{align}
\Delta t\e{sha} & = - 4 G(1+z_l) \l(1+\frac{m^2}{2\omega^2}\r) \\
 & \quad \times\int_{\mathbb{R}^2} \dd^2 \bs{\xi'} \Sigma(\bs{\xi'})\log\l( ||\bs{\xi}-\bs{\xi'}||\r)\,.
\end{align}
The same angular diameter distance ratio which appears in Eq.\,\eqref{eq:geom_time_delay} can be artificially pulled out 
\begin{align}
\Delta t\e{sha} = - \l( 1+\frac{m^2}{2\omega^2}\r) (1+z_l)\frac{D_s D_l}{D_{ls}} \psi(\bs{\theta}) \label{eq:Shap_time_delay}
\end{align}
to define the \textit{lensing potential} \cite{Kochanek:2004ua} 
\begin{align}
\psi(\bs{\theta}) = \frac{1}{\pi} \int_{\mathbb{R}^2} \dd^2\bs{\theta'} \kappa(\bs{\theta'}) \log|| \bs{\theta}- \bs{\theta'}|| + \hbox{const}
\end{align}
which is defined in terms of the \textit{convergence} $\kappa(\bs{\theta}) = \Sigma(\bs{\theta})/\Sigma\e{c}$ with the critical surface density defined as $
\Sigma\e{c} \equiv \frac{1}{4\pi G} \frac{D_s}{D_{ls} D_l}$ (see, for example \cite{Schneider:1992}). 

\subsection{Total time delay}
\label{sec:total_time-delay}

\noindent In this section, we express the total time delay for a massive graviton, which we will denote with a subscript $\rm{g}$, as in $\Delta t\e{g}$, and compare it to the time delay of a massless particle such as photons, for which quantities carry the subscript $\gamma$, as in $\Delta t_\gamma$.

\begin{figure}
\centering
    \includegraphics[width=0.95\columnwidth]{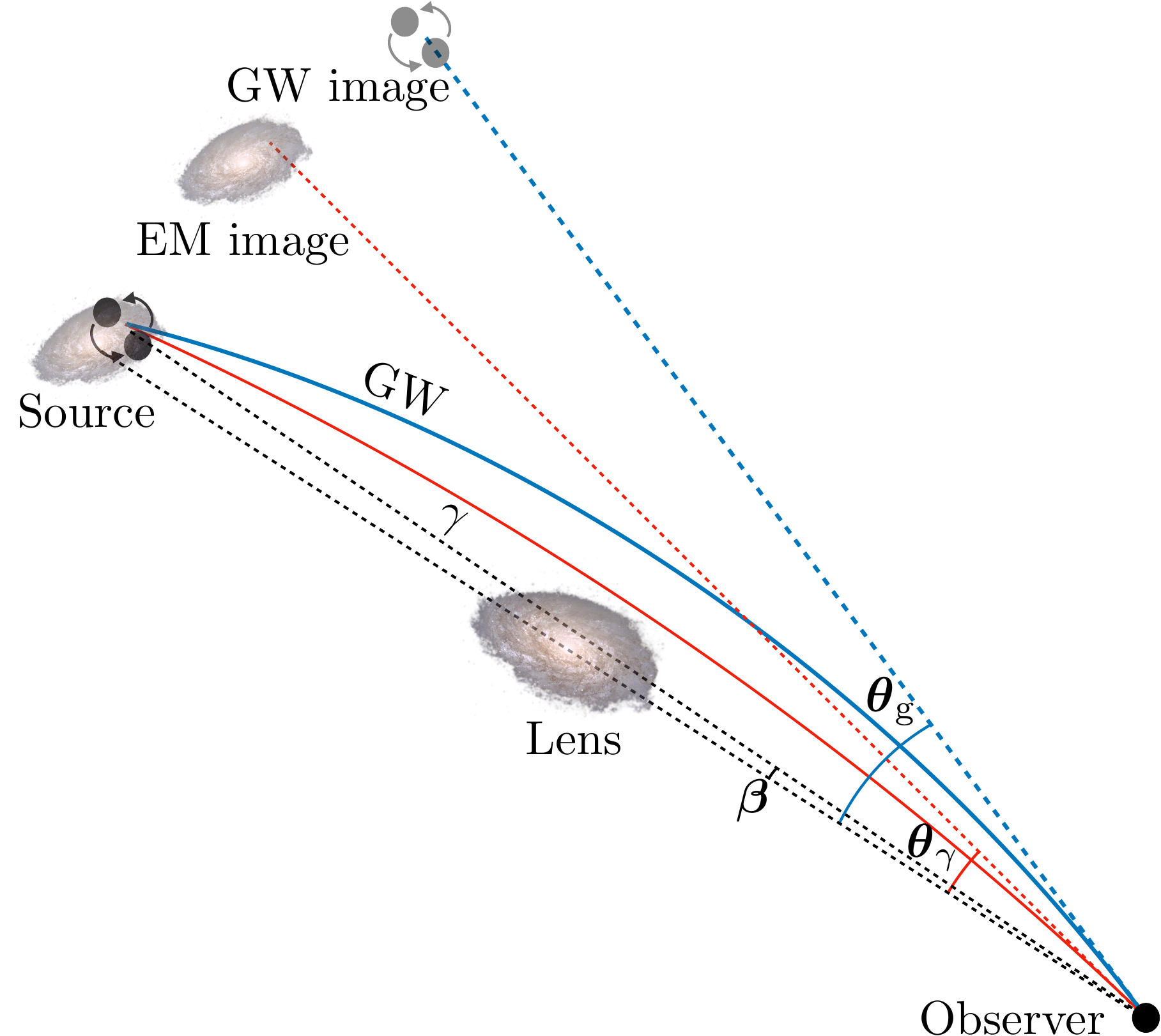}
    \caption{We show schematically the paths followed by the massive GW (in blue) which differs from the path followed by the massless photon ($\gamma$) in red. This results in different apparent images forming at different angles $\bs{\theta}\e{g}$ and $\bs{\theta}_\gamma$ for these signals, as found in Eq.\,\eqref{eq:images_GW_vs_EM}. The fact that the path for massive gravitons is a little bit longer geometrically turns out to be canceled by the fact that it also experiences less Shapiro time delay. This lucky cancellation results in the time delay between different images to differ only by the different group velocity between photons and massive gravitons as in Eq.\,\eqref{eq:model-independent delay}. We also show the source angle $\bs{\beta}$ which is the angle between the source and the optical axis, which connects the observer with a reference point in the lens.}
    \label{fig:Massive_graviton_lensing}
\end{figure}

The total time delay between images is given by the sum of the geometric and Shapiro time delays given by Eq.\,\eqref{eq:geom_time_delay} and Eq.\,\eqref{eq:Shap_time_delay}
\begin{align}
%\label{eq:total delay}
    \nonumber\Delta t\e{g} & = \left(1+\frac{m^2}{2 \omega^2}\right)  (1 + z_l)\frac{D_sD_l}{D_{ls}} \hat{\phi}(\bs{\theta}\e{g}^{\rm I},\bs{\beta}) \\
    & = \frac{\Delta t[\bs{\theta}\e{g}^{\rm I},\bs{\beta}]}{v\e{g}}  \,,\label{eq:graviton_time_delay}
\end{align}
where we have identified the GW group velocity as
\begin{align}
v\e{g} \equiv \frac{\p \omega}{\p k} \simeq 1 - \frac{m^2}{2\omega^2}\,,\label{eq:group_velocity}
\end{align}
and the standard time delay as a function of the image angle $\bs{\theta}$ and the source position $\bs{\beta}$
\begin{align}
\Delta t[\bs{\theta},\bs{\beta}] = (1 + z_l)\frac{D_sD_l}{D_{ls}}\hat{\phi}(\bs{\theta},\bs{\beta}) \,, \label{eq:standard_time_delay}
\end{align}
written in terms of the \textit{Fermat potential}
\begin{align}
\hat{\phi}(\bs{\theta},\bs{\beta}) = \left[\frac{(\bs{\theta} -\bs{\beta})^2}{2}- \psi(\bs{\theta})\right] \,. \label{eq:Fermat_potential}
\end{align}
Note that Eq.\,\eqref{eq:graviton_time_delay} confirms the intuition of \cite{Chung:2021rcu}. 
To make connection with the electromagnetic time delay, one must acknowledge that the images form at different angles $\bs{\theta}\e{g}^{\rm I}$ than in the massless case $\bs{\theta}_\gamma^{\rm I}$, as we have found in Sec.\,\ref{sec:scattering_angle}. This comes from the fact that massive gravitons are more deflected than massless photons such that the geometric time delay is longer. However, since they hit the lens plane with a larger impact parameter, they also experience less Shapiro time delay. Expanding $\bs{\theta}\e{g}^{\rm I}$ around $\bs{\theta}_\gamma^{\rm I}$ to first order in $m^2/\omega^2$, one obtains that the massive GW time delay $\Delta t\e{g}$ relates to their massless counterpart $\Delta t_\gamma$ as 
\begin{align}
\Delta t\e{g} = \frac{\Delta t_\gamma}{v\e{g}} + \frac{m^2}{\omega^2}(1 + z_l)\frac{D_sD_l}{D_{ls}} \bs{\theta}_\gamma^{\rm I} \cdot \l[(\bs{\theta}_\gamma^{\rm I}- \bs{\beta}) - \bs{\nabla}\psi(\bs{\theta}_\gamma^{\rm I})\r]\,. \label{eq:time_delay_extra_terms}
\end{align}
where the electromagnetic time delay $\Delta t_\gamma$ can be defined in terms of the standard time delay [Eq.\,\eqref{eq:standard_time_delay}]
\begin{align}
\Delta t_\gamma = \Delta t[\bs{\theta}_\gamma^{\rm I},\bs{\beta}]\,. \label{eq:Delta_t_gamma}
\end{align}
The experienced lensing reader might recognize the lensing equation in the square brackets of Eq.\,\eqref{eq:time_delay_extra_terms}, which tell that EM images form at positions $\bs{\theta}_\gamma^{\rm I}$, which extremize the Fermat potential
\begin{align}
\bs{\beta} = \bs{\theta}_\gamma^{\rm I} - \bs{\alpha}(\bs{\theta}_\gamma^{\rm I})
\end{align}
with $\bs{\alpha} (\bs{\theta}_\gamma^{\rm I}) =\bs{\nabla}\psi (\bs{\theta}_\gamma^{\rm I})$. Hence, the second term of Eq.\,\eqref{eq:time_delay_extra_terms} drops, and we are left with the very elegant result
\begin{empheq}[box=\fbox]{equation}
\Delta t\e{g} = \l( 1+\frac{m^2}{2\omega^2}\r)\Delta t_\gamma\,.\label{eq:model-independent delay}
\end{empheq}
Equation\,\eqref{eq:model-independent delay} is the main result of this article. It relates the time delay between different images in the GW and EM sector if GWs obey a massive dispersion relation, as in Eq.\,\eqref{eq:kmu_kmu}. The interpretation is clear, massive gravitons travel slower than massless photons over the path difference of different images. It is quite remarkable that even if they travel through different paths than massless particles, the amount of time gained geometrically is canceled by experiencing less Shapiro time delay. We depict this in Fig.\,\ref{fig:Massive_graviton_lensing}. Also note that while we computed the scattering angle for a pointlike lens in Sec.\,\ref{sec:scattering_angle}, this cancellation is completely general and holds for arbitrary lens models. This fact has been underappreciated in the literature \cite{Collett:2016dey}, where it was assumed that in modified gravity, GWs traveling at $c\e{g}<1$ also travel along the same paths, although this is generically incorrect as shown in Sec.\,\ref{sec:scattering_angle}. It had to be similarly assumed that the Shapiro time delay is unaffected by modified gravity. We showed here instead that if the GWs follow different paths than photons, the two contributions individually differ from the massless case and they cancel in the time delay.

We now discuss the observational prospects. Suppose that we observe a multimessenger event. In a \textit{golden} scenario, one can imagine observing a strongly lensed gravitational wave in coincidence with an electromagnetic counterpart. This would lead to a time-delay measurement in the electromagnetic domain ($\Delta t_\gamma$) and one in the gravitational wave ($\Delta t\e{g}$) domain. These should be related as in Eq.\eqref{eq:model-independent delay}. Rearranging this formula, the mass of the graviton can be expressed as
\begin{align}
m = \frac{\sqrt{2}\hbar\omega}{c^2} \sqrt{\frac{ \Delta t\e{g}}{\Delta t_\gamma}- 1}\,,
\end{align}
where we have temporarily introduced factors of $\hbar = 1 = c$. The advantage of the multimessenger measurement is quite clear. It does not require any modeling of the lens or of cosmology, which disappears in the comparison between the electromagnetic and GW time delay. This is quite remarkable, given that they follow different geodesics and experience different local Shapiro time delays. We may even get a different time delay for each frequency $\Delta t\e{g}(\omega)$ if we can identify corresponding crests of the GW for the different signals. 

Note that the advantage with respect to dispersion relation constraints on the speed of the low versus high frequencies of the GW, is that it does not require us to estimate the GW phase at the source, which requires a waveform model in the presence of a massive graviton. Here instead, one can identify the merger time and the peaks of the GW in a model independent way. This allows for a fully model-independent constraint on the graviton mass. Hereafter, we evaluate whether this constraint is competitive. 

In the null mass case, the time delays are equal, such that $ t\equiv \Delta t\e{g}- \Delta t_\gamma =0$ up to some uncertainty $\sigma_{\Delta t} = \sqrt{\sigma_{\Delta t\e{g}}^2 + \sigma_{\Delta t_\gamma}^2}$. For one time-delay measurement, we take two time-of-arrival measurements, whose errors $\sigma_t$ are added in quadrature. \footnote{Hence $\sigma_{\Delta t} = \sqrt{2}\sigma_{t}$ for both GW and electromagnetic signals.}
The $95\%$ confidence limit on $m$ reads
\begin{align}
m <  \frac{2 \hbar \omega}{c^2} \sqrt{\frac{\sqrt{\sigma_{\Delta t\e{g}}^2 + \sigma_{\Delta t_\gamma}^2}}{\Delta t_\gamma}}\,.
\end{align}
Note that the individual contributions of geometric and Shapiro effects to the total time delay in the denominator are degenerate. The degeneracy is broken by the details of the lens model, which drops out of this constraint. As such, the constraint holds regardless of the precise cancellation between the Shapiro and geometric time delays.

The uncertainty on the merger time of the gravitational wave is determined by how well one can resolve the waveform. 

We adopt a time-delay error in the GW signal of $10^{-6}$ for the LIGO band, and $0.1$s for LISA, which is predicted to be feasible for high signal to noise ratio events \cite{Marsat:2020rtl, Sharma:2024sfb}. This distinction is not apparent in the plot as the term in the square root is dominated by the EM time-delay errors.

\begin{figure}
\includegraphics[width=1.0\linewidth]{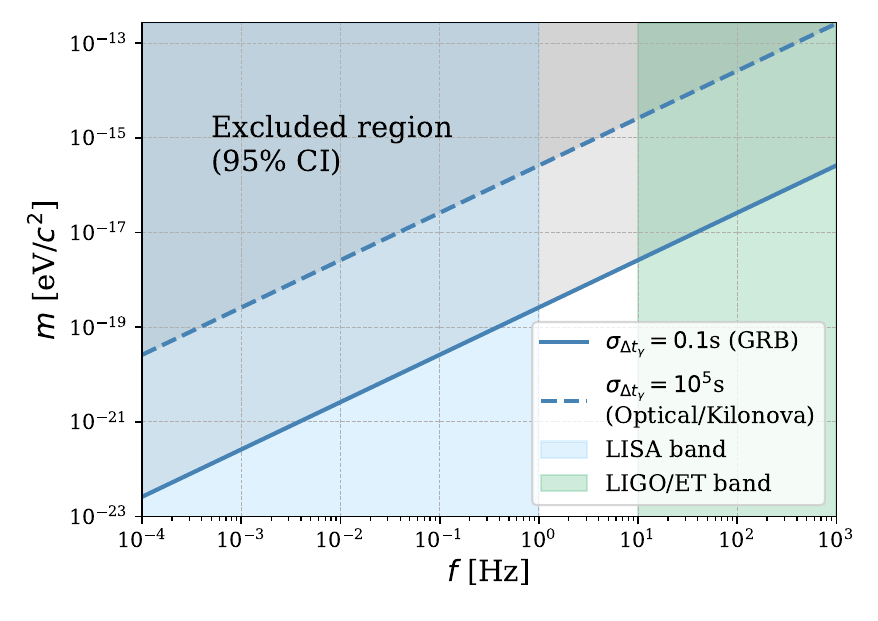}
    \caption{Upper bound on the graviton mass as a function of gravitational wave frequency. The solid line corresponds to an optimistic scenario with an error on EM time-delay measurements of $0.1$s, while the dashed line corresponds to an error of $10^5$s. The gray areas above the lines correspond to the values of $m$ excluded at $95\%$ confidence. The area shaded in blue corresponds to the LISA frequency band, while the green area corresponds to ground-based detector bands. Constraints shown are for a galactic cluster lens.
    }
    \label{fig:error on mg}
\end{figure}

For the uncertainty on the electromagnetic time delay, we consider two scenarios. In an optimistic case, the time delay can be determined to subsecond precision $\sigma_{\Delta t_\gamma} = 0.1$s, as might be the case for a short gamma ray burst \cite{LIGOScientific:2017ync}. A more pessimistic EM time-delay uncertainty could be $\sigma_{\Delta t_\gamma} = 10^{5}$ s (similar to the typical kilonova time-delay uncertainty \cite{Breschi:2021tbm}). We plot the $2\sigma$ upper limit on $m$ as a function of frequency $f$ (where $f= \omega/(2\pi)$) for these two scenarios for a signal with $\Delta t_\gamma = 1000$ days in Fig.\,\ref{fig:error on mg}. Note that this time delay is typical for galactic cluster lenses and that galaxy lenses would typically give lower time delays and, hence, weaker constraints on the graviton mass. The tightest constraint shown is $m< 3\cdot 10^{-23}$eV/c$^2$ for one measurement of $\Delta t_\gamma$ and $\Delta t\e{g}$ (i.e. two images detected) in the LISA band, with $f = 10^{-4}$Hz lensed by a cluster. If we were to detect a quadruply imaged signal, we would obtain three independent measurements of both time delays which implies $m < 2 \cdot 10^{-23}$eV/c$^2$. This constraint is comparable to current dispersion relation bounds \cite{LIGOScientific:2021sio} with the advantage of being independent of the lens, waveform, and cosmological models. Additionally, one could in principle measure the time delay at different frequencies if one can identify precisely waveform features of the different images, which would provide additional independent measurements of $\Delta t\e{g}$.

As mentioned above, we are working in the geometric optics approximation, which may break down when the wavelength of the GW becomes smaller or comparable to the Schwarzschild radius of the lens, $\lambda \approx R\e{S}$. This may happen at low frequencies for certain galaxy-scale lenses, in which case the formalism that we use is unsuitable, and one should resort to wave optics.\footnote{An example case where geometric optics breaks down occurs at frequencies around $10^{-7}$Hz, which is much lower than LISA merger frequency. At this frequency the GW wavelength is $\lambda \approx 10^{15}$m, comparable to a Schwarzschild radius of a galaxy similar to the Milky Way.} For galactic cluster lenses, with $R\e{S}$ of the order of light years, this is at least 4 orders of magnitude larger than a GW wavelength at the lowest frequency we consider ($10^{-4}$Hz, which corresponds to roughly $10^{12}$ m). In this scenario, the quoted constraints in Fig.\,\ref{fig:error on mg} are safely within the geometric optics regime. 

\section{Magnification constraint}\label{sec:magnification_constraint}

\noindent In this section, we derive the amplification factor of gravitational waves in the presence of a nonzero mass in the dispersion relation [Eq.\,\eqref{eq:kmu_kmu}]. We first show that the Kirchhoff diffraction integral is unaffected by the graviton mass in Sec.\,\ref{subsec:Kirchhoff_theorem}. We then compute the amplification factor in Sec.\,\ref{subsec:amplification_factor}. Finally, we show in Sec.\,\ref{sec:magnification_derivation} that the comparison of EM versus GW magnification leads to worse constraints on the graviton mass than the time delay technique.

\subsection{Kirchhoff's theorem}\label{subsec:Kirchhoff_theorem}
\begin{figure}
    \centering
    \includegraphics[width=1.0\columnwidth]
    {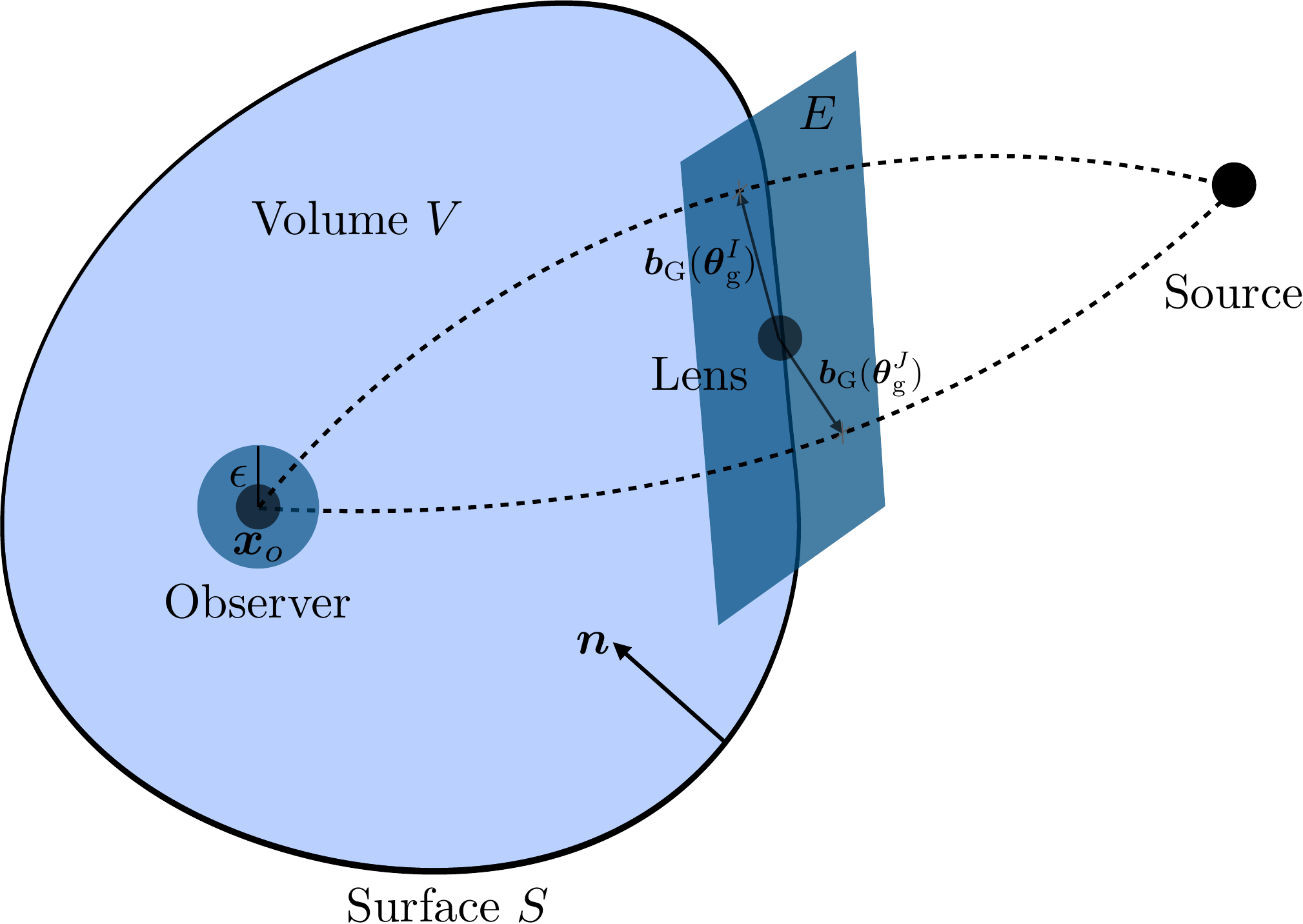}
    \caption{The Kirchhoff theorem setup. We consider a volume $V$ in blue bounded by a surface $S$ to evaluate the wave amplitude at point $\bs{x}_o$, which corresponds to the observer. The source is located outside the volume $V$. We also depict a sphere of small radius $\epsilon$ around the observer (the excluded region in our calculation) and the vector $\bs{n}$, which is normal to the surface $S$. Additionally, though the Kirchhoff theorem is general, we show the positions of a source, observer and lens together with the paths followed by the geometric optics images which form at impact parameters $\bs{b}\e{G}(\bs{\theta}^{\rm{I}}\e{g})$ and $\bs{b}\e{G}(\bs{\theta}^{\rm{J}}\e{g})$ in the lens plane to facilitate the connection between our derivation of the theorem and its use in the context of lensing.
    }
    \label{fig:kirchhoff-final}
\end{figure}
\noindent To evaluate how the amplification factor of a lensed gravitational wave differs in the presence of a graviton mass, we start by demonstrating that Kirchhoff's theorem holds unchanged. This is true despite the modified dispersion relation. This theorem allows one to express the amplitude of the wave at a point in terms of the wave evaluated on a closed surface. This is useful when one knows the amplitude of a wave on that surface but not at the point in question, because of intervening matter between the source and the observer, as is the case in a gravitational lensing situation. Following the derivation in Chap. 8.3 of Ref.\,\cite{Born:1999ory}, we assume that the GW is well described by a scalar wave, which we write in Fourier space\footnote{Note that we can treat GWs as scalar waves, since to lowest order in geometric optics, the polarization is parallel transported along the geodesic \cite{Misner:1973prb}, but see \cite{Cusin:2019rmt,Dalang:2021qhu} for beyond geometric optics effects.} 
\begin{align}
U(\bs{x}, t) = \int_{\mathbb{R}} \frac{\dd \omega}{\sqrt{2\pi}} \t{U}(\omega, \bs{x})\hbox{e}^{-\mathrm{i}\omega t} \, ,
\end{align}
and satisfies a vacuum Klein-Gordon equation $(\Box 
-m^2)U = 0 $ inside a volume $V$.  This implies that each Fourier mode satisfies
\begin{align}
\label{eq:KG}
(\bs{\nabla}^2 - m^2 + \omega^2)\t{U} = 0
\end{align}
on Minkowski space. Suppose that $\t{U}'$ is another solution of Eq.\,\eqref{eq:KG}. Following the setup showin in Fig.\, \ref{fig:kirchhoff-final}, if $\bs{n}$ is the inward normal to a closed surface $S$, and $\t{U}$ and $\t{U}'$'s first and second partial derivatives are continuous within and on $S$, we can apply Green's theorem over the enclosed volume $V$
\begin{align}
\label{eq:green's theorem}
    \int_V (\t{U} \bs{\nabla}^2\t{U}'-\t{U}'\bs{\nabla}^2\t{U}) \dd V = -\oint_S \dd^2\bs{n} \cdot \left(\t{U} \bs{\nabla} \t{U}'-\t{U}'\bs{\nabla} \t{U}\right)  \, .
\end{align}
Substituting Eq.\,\eqref{eq:KG} into the left-hand side, the integral vanishes in the same way it does in the massless case, because scalars commute. We are left with
\begin{align}
    0 = \oint_S \dd^2\bs{n} \cdot \left(\t{U} \bs{\nabla} \t{U}'-\t{U}'\bs{\nabla} \t{U}\right) \, .
\end{align}
To evaluate the field at a point $\bs{x}_o$ within the volume, we take $U' = \exp{\left(iks\right)/s}$, where $s$ is the distance from $\bs{x}_o$ and $k = \sqrt{\omega^2-m^2}$ is the wavenumber. One can check that $U'$ indeed satisfies Eq.\,\eqref{eq:KG}. However, $U'$ has a singularity at $s=0$ on the point $\bs{x}_o$ which needs to be excluded from the closed surface integral. One can close the surface integral on a small sphere of radius $\epsilon$ centered on $\bs{x}_o$ and take the $\epsilon \to 0$ limit. The only surviving term is $-4\pi U(\bs{x}_o)$, which leads to the \textit{Kirchhoff} diffraction integral
\begin{align}
     \t{U}(\omega, \bs{x}_o) = \frac{1}{4\pi} \oint_S \dd^2\bs{n} \cdot \l[ \t{U} \bs{\nabla}\l(\frac{e^{\mathrm{i}ks}}{s}\r) - \frac{e^{\mathrm{i}ks}}{s} \bs{\nabla}\t{U} \r] \,.
\end{align}
This result is identical to the massless case, although here $k = \sqrt{\omega^2 - m^2} \neq \omega$ due to the modified dispersion relation. It allows to express the value of a field $\t{U}$ at a certain position $\bs{x}_o$ in terms of the same field integrated over a closed surface. Again, this is useful if one knows the value of that field on the surface, but not at $\bs{x}_o$, because of intervening matter along the line of sight, as may be the case in gravitational lensing scenarios. 

\subsection{Amplification factor}\label{subsec:amplification_factor}

\noindent In this section we derive the amplification factor from the Kirchhoff diffraction integral. The amplification factor is defined as the ratio between the lensed waveform $\t{h}(\omega,\bs{x}_o)$ and unlensed waveform $\t{h}_{\rm{nolens}}(\omega,\bs{x}_o)$ for each image
\begin{align}
F\e{g}(\omega) \equiv \frac{\t{h}(\omega,\bs{x}_o)}{\t{h}_{\rm{nolens}}(\omega,\bs{x}_o)} \,.
\end{align}
We consider the lensing situation depicted in Fig.\,\ref{fig:lensing_configuration}. We make a thin lens approximation and consider a plane  $E$, which is sufficiently far from the lens such that spacetime can be considered to be Minkowski space between the observer and that plane. At the end of the calculation, we take the limit in which the distance between that plane and the lens is much smaller than the distance between the lens and the observer, such that on Fig.\,\ref{fig:lensing_configuration}, the lens and the plane $E$ appear to coincide. This allows us to use the Kirchhoff diffraction integral to write the amplitude of the wave at the observer in terms of an integral over the plane $E$, as 
\begin{align}
\t{h}(\omega, \bs{x}_o) = \frac{1}{4\pi} \int_E \dd^2 \bs{n} \cdot \l[ \t{h} \bs{\nabla} \l( \frac{e^{\mathrm{i}k d_l}}{d_l}\r) - \frac{e^{\mathrm{i}kd_l}}{d_l} \bs{\nabla}\t{h}\r]\,.
\end{align}
We also consider that on that plane, the Shapiro time delay was already effective such that the wave at the lens, with impact parameter $\bs{b}$, can be written as
\begin{align}
\t{h}(\omega, \bs{b}) = \t{H}(\omega,\bs{b}) \exp \l\{ \mathrm{i} k\l(\Delta t\e{g}[\bs{\theta}(\bs{b}),\bs{\beta}] + d_s - d_l\r)\r\}
\end{align}
which is written in terms of an amplitude $\t{H}(\omega,\bs{b})$ on the lens plane and $\Delta t\e{g}(\bs{\theta}(\bs{b}),\bs{\beta})$ which captures the time (or alternatively distance) delay due to the lens that we have calculated in Eq.\,\eqref{eq:graviton_time_delay}. In the exponent, $(ik)$ multiplies an effective traveled distance between the source and the lens plane at impact parameter $\bs{b}$. Under the assumptions that it is enough to evaluate the slowly varying amplitude on the geometric optics path defined by the impact parameter on the lens plane $\bs{b}\e{G}$, that the scattering angle is small, and that $k d_l \gg 1$, we get
\begin{align}
\t{h}(\omega, \bs{x}_o)  = \frac{\mathrm{i} k }{2\pi} \frac{\t{H}(\omega,\bs{b}\e{G})}{d_l} \int_E \dd^2 \bs{b} e^{\mathrm{i} k (\Delta t\e{g}[\bs{\theta}(\bs{b}),\bs{\beta}] + d_s )}\,.
\end{align}
By noticing that the amplitude of the wave is inversely proportional to the comoving distance, one can check that
\begin{align}
\t{H}(\omega,\bs{b}\e{G}) = \frac{\t{H}(\omega,\bs{\eta})}{d_{ls}} = \frac{\t{H}_{\rm{no lens}}(\omega,\bs{x}_o) d_s }{d_{ls}}
\end{align}
and that $\t{h}_{\rm{no lens}} (\omega,\bs{x}_o) = \t{H}_{\rm{no lens}}(\omega,\bs{x}_o) e^{\mathrm{i} k d_s}$. We then find
\begin{align}
\label{eq:F_g}
F\e{g} = \frac{\mathrm{i} k D_s D_l}{D_{ls}} \int_{\mathbb{R}^2} \dd^2 \bs{\theta} \exp \l(\mathrm{i} k \Delta t\e{g}[\bs{\theta} , \bs{\beta}]\r) \,,
\end{align}
where we have also changed the surface integral to an angular integral, using $\dd^2\bs{b} = D_l^2 \dd^2\bs{\theta}$ and used the relation between comoving distances and angular diameter distances and the fact that the volume between the lens plane and the observer has been assumed to be flat spacetime. To linear order in $m^2/\omega^2$, we find that
\begin{align}
\label{eq:FgFgamma}
F\e{g} = \l(1-\frac{m^2}{2\omega^2} \r) F_\gamma\,, 
\end{align}
with the standard EM amplification factor $F_\gamma$ being \cite{Schneider:1992}
\begin{align}
F_\gamma = \frac{ \mathrm{i} \omega}{2\pi} \frac{ D_s D_l}{D_{ls}} \int_{\mathbb{R}^2} \dd^2 \bs{\theta} \exp \l(\mathrm{i} \omega \Delta t[\bs{\theta},\bs{\beta}]\r) \,, \label{eq:standard_amplification_factor}
\end{align}
where the mass has disappeared from the exponent in the integral and $\Delta t[\bs{\theta},\bs{\beta}]$ indicates the standard time delay, defined in Eq.\,\eqref{eq:standard_time_delay}. This result, which we derived from first principles, differs from the phenomenological approach in \cite{Chung:2021rcu}, which has a factor of $m^2/\omega^2$ in the exponent and a sign difference in the prefactor. Since the mass correction enters as a ratio $m^2/\omega^2$ as seen in Eq.\,\eqref{eq:FgFgamma}, the amplification factor mass correction is frequency dependent. This feature can be used to extract the mass of the graviton at relatively low frequencies, as discussed in \cite{Geng:2025}.

Equation\,\eqref{eq:FgFgamma} suggests that one can use the comparison of the GW and EM amplifications to constrain the graviton mass.\footnote{The EM amplification is not directly accessible because one observes fluxes instead of the amplitude of the EM wave, which means one has to resort to magnifications, as discussed in the next section.} Note that this assumes that the source angles are the same, i.e.\,$\bs{\beta}\e{g}= \bs{\beta}_\gamma$. This might be the case for an EM transient emitted along the GW signal.
%This problem, however, is overshadowed by a more important one. 
For an extended source, such as a galaxy, it becomes unclear where the GW signal originates from within the host galaxy, such that in general, $\bs{\beta}\e{g} \neq \bs{\beta}_\gamma$. The difference in amplification factor between a GW being emitted from one side of the galaxy and the other might be larger than the difference generated by the mass of the graviton. This would of course spoil the test. In the following, we show that even in the best scenario, this method to constrain the graviton mass is weaker and less practical than the time-delay technique, as it requires assumptions about the lens model and cosmology

\subsection{Magnification}\label{sec:magnification_derivation}

\noindent In this section, we show that the constraint that can be set on the graviton mass from the magnifications is weaker than the constraint that can be obtained from the time delay. We first clarify how the magnification relates to the amplification factor.

The amplification factor results in real space as an amplitude and phase shift. For a monochromatic signal, the lensed waveform is magnified according to
\begin{align}
h(t,\bs{x}_o) = \sqrt{\mu\e{g}} h\e{nolens}(t,\bs{x}_o)
\end{align}
where $\mu\e{g} \equiv |F\e{g}|^2$ denotes the GW magnification. Hence, the observed luminosity distance extracted from the waveform $D\e{L}$ relates to the background\footnote{For example, the luminosity distance that one would calculate in a perfect FLRW universe.} luminosity distance $\bar{D}\e{L}(z)$ to the GW source that one would compute given an observed redshift $z$ and cosmological model as follows
\begin{align}
D\e{L} = \bar{D}\e{L}(z)/\sqrt{\mu\e{g}}\,, \label{eq:mag_g}
\end{align}
which is a well-known result. In comparison, electromagnetic waves have their observed flux $\Phi$ magnified as
\begin{align}
\Phi = \mu_\gamma \Phi\e{nolens}\,. \label{eq:mag_gamma}
\end{align}
Here, $\Phi$ is proportional to the square of the amplitude of the EM wave. The GW magnification relates to the EM magnification $\mu_\gamma = |F_\gamma|^2$ as %|\Phi_\gamma|^2$ 
\begin{align}
\mu\e{g} = \l(1- \frac{m^2}{\omega^2}\r) \mu_\gamma\,.\label{eq:magnifications}
\end{align}
Equation\,\eqref{eq:magnifications} suggests that one can use magnifications to constrain the graviton mass. We consider again the \textit{golden} scenario, where one observes a lensed GW and the lensed host galaxy, both of which are magnified according to Eqs.\eqref{eq:mag_g} and \eqref{eq:mag_gamma} with $\mu^{\rm{I}}_{g ,\gamma}$ with $\rm{I}\in \{1,\dots, N \}$ for $N$ images. While we have $2N$ measurements ($N$ $h^{\rm{I}}$ for the GW images and $N$ $\Phi^{\rm{I}}$ for the EM images), we need extra information to break the degeneracy between $h\e{nolens}$, $\Phi\e{nolens}$ and the magnifications $\mu\e{g}^{\rm{I}}$ and $\mu_\gamma^{\rm{I}}$ which are unobservable and represent $2N + 2$ unknowns. The EM magnification can be estimated from a lens model \cite{Schneider:2013sxa}, for which current error bars are of the order of $20\%$ due to the mass-sheet degeneracy \cite{Birrer:2020tax, Hannuksela:2020xor}. The GW magnification can be extracted from the luminosity distance encoded in the waveform, the source redshift and a cosmological model, which allow one to compute $\bar{D}\e{L}(z)$.
In this case the current error bars can be of the order of $30\%$, driven by errors in the observed luminosity distance which are largely due to its degeneracy with inclination. Under these assumptions, the inferred magnifications of each image can be used to extract the graviton mass which reads
\begin{align}
m= \frac{\hbar \omega}{c^2} \sqrt{1- \frac{\mu\e{g}^{\rm{I}}}{\mu_\gamma^{\rm{I}}}} \,, \quad \forall\, \rm{I} \in\{1,\dots,N\}
\end{align}
where we have written factors of $\hbar = 1=c$ explicitly. In an analogous way as for the time delay, the 95$\%$ confidence upper limit on the graviton mass reads
\begin{align}
\nonumber m & < \frac{\sqrt{2}\hbar \omega}{c^2} \sqrt[4]{\l(\frac{\sigma_{\mu\e{g}}}{\mu\e{g}}\r)^2 +\l(\frac{\sigma_{\mu_\gamma}}{\mu_\gamma}\r)^2 } \\
& = 6 \cdot 10^{-19}\hbox{eV}/c^2 \l(\frac{f}{10^{-4} \hbox{Hz}}\r)\sqrt[4]{\l(\frac{\sigma_{\mu\e{g}}}{\mu\e{g}}\r)^2 +\l(\frac{\sigma_{\mu_\gamma}}{\mu_\gamma}\r)^2 }\,.
\end{align}
Even under the most optimistic scenario -- assuming a perfect lens model, known cosmology, and minimal GW luminosity distance error (e.g. from a particularly loud event or broken distance-inclination degeneracy) -- yielding relative error bars on the magnifications of around $1\%$, this constraint remains at least 3 orders of magnitude weaker than the one from the time delay. It also requires modeling the lens to extract the EM magnifications $\mu_\gamma^{\rm{I}}$ and to assume a cosmological model to extract the GW magnification $\mu^{\rm{I}}_g$. Furthermore, we had to assume that the source angle in the EM and GW are the same, which may not be a good assumption if the EM source is extended, like a galaxy. Note that if the theory allows for extra polarizations, some power may be lost from the spin-2 channels. This could lead to biases in the GW distance and magnification if this effect is not properly taken into account, which represents further complications. However, testing for extra polarizations can be done in the absence of lensing \cite{LIGOScientific:2021sio}. In all regards, this constraint is weaker than the constraint coming from the time delay.

\section{Conclusion}\label{sec:conclusion}

\noindent In this work, we have investigated the effects of a massive graviton on a lensed gravitational wave, and the potential for a lensed multimessenger event to constrain the mass of the graviton. After illustrating how a massive graviton affects the dispersion relation, we investigated three aspects that change in the presence of a massive graviton, namely, geodesics, the time delays between different images, and the magnification of the signals due to a gravitational lens. We computed the first-order corrections to these quantities in powers of $m^2/\omega^2$.

First, we solved the geodesic equation and showed that the scattering angle is affected by the mass of the graviton. Note that the scattering angle generally differs from the massless case for waves traveling at subluminal speeds.

Starting from the dispersion relation, we derived an expression for the time delay between different lensed signals for a massive graviton, which differs from the massless case by a factor of $1-m^2/(2\omega^2)$. Accounting for the different scattering angles of massive gravitons with respect to massless photons led to two extra terms which turn out to cancel. This is because while the massive gravitons travel a little bit longer geometrically than photons, because of their larger scattering angle, this is compensated by experiencing less time dilation from the lens. This is due to the impact parameter being larger than for massless photons, as we illustrate in Fig.\,\ref{fig:Massive_graviton_lensing}. This cancellation makes the difference in time delay between photons and massive gravitons arise solely because of the different group velocity of the gravitons with respect to massless photons, as may be understood from Eq.\,\eqref{eq:model-independent delay}. We argued that we could constrain the mass of the graviton by comparing GW and EM time delays between different images. This effectively cancels the contributions that depend on cosmology and the lens model, as should be clear from Eq.\,\eqref{eq:model-independent delay}, such that a fully model-independent constraint can be imposed on the mass of the graviton. We find that for a merger in the milli-Hertz which is relevant for LISA, the constraint can reach $m < 3 \cdot 10^{-23}$eV/c$^2$ when lensed by a cluster-scale lens. Note that, additionally, the constraint is independent of the waveform model, which is not the case for the dispersion constraints.

We then focused on the magnification of the GWs. We showed that Kirchhoff's diffraction formula is valid for the massive case and used it to compute the amplification factor. We find that the GW amplification factor differs from what was previously used in the literature \cite{Chung:2021rcu}. We then studied how the amplitude of the signal can be used to set a constraint on the graviton mass. We find that it is difficult to use the magnifications to this end, since it generally requires a lens model and a cosmological model to set tight constraints on the graviton mass. Additionally, if one uses an extended source such as a galaxy or a quasar to observe the EM magnification of images, the difference induced with respect to the GW magnification may spoil the graviton mass constraint because magnification is very sensitive to the source position. 

Finally, microlensing by stars in the lens could in principle affect the EM and GW signals causing a discrepancy in both magnification and time delays. This holds provided that the geometric optics approximation remains valid, which is less obvious for GWs than it is for EM waves, given the wavelengths involved. In such a scenario, since the geodesics followed by massive gravitons and light are slightly different, there is a possibility that one of the two gets micromagnified but not the other. Such a signal could be used as a smoking gun signature of a nonzero graviton mass, and we leave this possibility to future work. As for the time delay, microlensing of the EM signal effectively adds an extra delay term to the right-hand side of Eq.\,\eqref{eq:Delta_t_gamma}. This is, however, of order $10^{-5}$ s \cite{Lewis:2020asm, Meena:2025cdo}, much smaller than both $\Delta t_\gamma$ and $\sigma_{\Delta t_\gamma}$, meaning that our estimate on the graviton mass constraints would remain unaffected. Note that a graviton charge would be partially degenerate with a graviton mass, as discussed in \cite{Nair:2024xdb}.

Overall, we have shown that a single multimessenger lensed event can probe the graviton mass in a fully model-independent fashion and constrain $m< 3 \cdot 10^{-23}$eV/c$^{2}$ using the time delay between different images. 

We have mentioned that lensed gravitational wave signals are expected to be rare, with only one observed multiply imaged signal every $\sim 1500$ detections. While the current number of GW detections and the absence of comprehensive all-sky coverage in the EM domain make the detection of a golden event currently unlikely, the methods described in this paper may be used in the foreseeable future. Over the coming decade, gravitational-wave astronomy will see significant advancements with the advent of LISA, Einstein Telescope, and Cosmic Explorer. The latter two in particular forecast detections in the thousands \cite{Maggiore:2019uih, Evans:2021gyd}, and with new surveys such as LSST monitoring large portions of the sky, there is a chance to observe a multimessenger strongly lensed event, which will enable a fully model-independent probe of the graviton mass.

\section{Acknowledgements}
\noindent We thank Wolfgang Enzi, Ana Sainz de Murieta, Tian Li, Ian Harry, Gareth Davies, Luke Weisenbach, and Adrian Ka-Wai Chung for interesting discussions. Additionally, it is a pleasure to thank Tom Collett and Martin Millon for useful comments on a preliminary version of this article. E.C., C.D.\,and T.B.\,are supported by ERC Starting Grant SHADE (Grant No.\,StG 949572). T.B.\,is further supported by a Royal Society University Research Fellowship (Grant No.\,URF\textbackslash R\textbackslash 231006). For the purpose of open access, the authors have applied a Creative Commons Attribution (CC BY) licence to any Author Accepted Manuscript version arising. 

\section{Data Availability}
\noindent No data were created or analyzed in this study.
\bibliographystyle{apsrev4-2}
\bibliography{main}

\end{document}